\documentclass[preprint,showpacs,preprintnumbers,amsmath,amssymb,aps]{revtex4}

\usepackage{graphicx}% Include figure files
\usepackage{dcolumn}% Align table columns on decimal point
\usepackage{bm}% bold math
\usepackage{amssymb}
\usepackage{epstopdf}
\usepackage{times}% correct fonts

\begin{document}

\title{High yield fusion in a Staged Z-pinch}

\author{H. U. Rahman, F. J. Wessel, N. Rostoker}
\affiliation{Department of Physics and Astronomy, University of California, Irvine, California 92697}

\author{P. Ney}
\affiliation{Mount San Jacinto College, Menifee, California}

%\date{\today}

\begin{abstract}

We simulate  fusion in a Z-pinch; where  the  load is
a xenon-plasma liner imploding onto a
deuterium-tritium plasma target and the driver is  a 2 MJ, 17 MA, 95 ns risetime pulser.  
The implosion system is modeled using  the dynamic, 2-1/2
D, radiation-MHD code, MACH2. During implosion 
a shock forms in the Xe liner,  transporting current and
energy radially inward. 
After collision with the DT,  a secondary shock forms 
pre-heating the DT to several hundred eV.  
Adiabatic compression leads subsequently to 
a fusion burn,
as the target is 
surrounded by a flux-compressed, 
intense, azimuthal-magnetic field. 
The intense-magnetic field confines  fusion
$\alpha$-particles, providing  an additional
source of ion heating that leads to  target ignition. 
The target remains stable  up to the time of
ignition. 
Predictions are for a neutron yield of $3.0\times 10^{19}$ 
and a thermonuclear energy of  
84 MJ, that is, 42 times greater than the 
initial, capacitor-stored energy. 
\end{abstract}

\maketitle

\bigskip
\noindent

\begin{center}
{\bf I. Introduction }
\end{center}

The simple Z-pinch is a cylindrical plasma column
that implodes to the axis of symmetry when  
subjected to a  large, sustained-current pulse.
A typical Z pinch load is constructed from
a wire-array, foil, plasma jet, or gas-puff, 
or combinations thereof. 
When driven by a modern, low-inductance,
high-voltage, pulse-power circuit, the pinch-current pulse
can reach many mega-amperes and the delivered power
100's of terawatts. 
Z-pinch plasmas with
keV temperatures and  near-solid densities
are produced routinely. 
The Z Facility at the Sandia National Laboratory 
is perhaps the best 
example: it produces a 20 MA, 
100 ns rise time, 100 TW current pulse and a  Z-pinch that 
can radiate  mega-joules of X-ray energy in a few nanosecond pulse. 
Such plasmas are of great scientific and technical interest,  for example in studies related to  fusion, 
atomic physics, laboratory astrophysics, etc. \cite{Matzen:1997,  Ryutov:2000, Haines:2000, Coverdale:2007a}.

The Z-pinch dynamics consists of three phases: implosion, stagnation, and disassembly. 
The implosion phase is when the discharge-current builds 
and  the pinch is driven radially inward by  
the $\vec J_z\times \vec B_{\theta}$ force, where $\vec J_z$ is the axial, 
plasma-current density and $\vec B_{\theta}$ is the azimuthal, self-magnetic field. 
At stagnation the pinch is confined 
briefly, typically for a few nanoseconds when the radial motion
of the pinch has ceased, or nearly so,  
and the plasma is compressed to a high-energy-density. 
Generally, the imploded mass and the 
initial-pinch radius are adjusted so 
that stagnation occurs  after the current maximum.  

At stagnation the implosion-kinetic energy  and the inductive energy,
stored local to the pinch, is rapidly converted  
into plasma-internal energy. Shock heating
is important. The small radius of 
the pinch plasma at stagnation also increases its electrical resistance, 
enhancing the energy deposited by  Ohmic heating. 
MHD instabilities occur in this phase: typically, for example 
the (m = 0) sausage instability or the (m = 1)  kink instability.

A fusion burn will result if the Z-pinch remains stable  
for a sufficiently long time,
while the required high temperature and density are sustained. 
Ignition is possible, if the fusion products are sufficiently well-confined.
This will occur if the azimuthal-magnetic field 
is sufficiently intense that   fusion $\alpha$-particles are confined, 
that is, $\rho_{\alpha} <<$ R$_{pinch}$, where 
$\rho_{\alpha}$ is the $\alpha$-particle 
gyro-radius and R$_{pinch}$ is the compressed pinch radius. 
Following stagnation the pinch disintegrates rapidly,
due to the rebound in plasma pressure and the accumulated 
effect of instabilities.

Z pinches are susceptible to the Rayleigh-Taylor (RT) 
instability during implosion. Many techniques 
have been developed to control the effect of the RT instability, 
all  directed  toward
maximizing the accumulated-pinch energy. 
The most common techniques consist of 
altering the load configuration to provide a more
uniform,  initial-mass distribution, or reducing the time 
needed to obtain a uniform, highly-conducting 
plasma at current
initiation. Other  approaches involve 
decreasing the risetime of the current pulse and 
using concentric, multi-layer mass distributions.

The gas-puff Z-pinch was developed in the 1970's 
as a stable alternative to the more widely used wire array Z-pinches
\cite{Shiloh:1978} and has demonstrated 
a surprisingly large range of scalability; having been  implemented on 
short- and long-implosion time generators, with risetimes, 
$\tau_{1/4} \sim 0.1 - 1~\mu$s
and load currents, I$_{load} \sim$ 0.1 - 10's of MA.
Gas-puff pinches have also been configured  to study 
gas mixtures. Gas-mixture Z-pinches have demonstrated a unique
ability to produce a  higher-energy radiated spectrum 
and higher  X-ray yield, than  a Z-pinch of either 
gas imploded separately \cite{Bailey:1986, Chang:1991, 
DeGroot:1997a, Chaikovsky:1997, Levine:2006, Qi:2008}. 
Multi-layer gas-puff implosions have also produced 
better results than single
layer, or uniform-fill Z-pinch. 

The improvements observed for gas mixtures and  multi-shell 
implosions suggests  that 
there is a complex interplay of shock-driven compressional  heating, 
current-diffusion, flux-compression, and radiation-transport at work, for
which further analysis will provide deeper insights 
\cite{Rostoker:1978, Velikovich:1996, Glazyrin:1997, DeGroot:1997b}. 
Staging the implosion, to optimize these 
dynamical processes, is expected to have specific benefits for fusion 
 \cite{Rahman:1989, Rahman:1995, Rahman:2004, Ney:2001}. 
A gas-puff mixture of deuterium and argon was 
tested  recently, with a reported  neutron yield of,  
$Y \sim 3.7-3.9\times 10^{13}$  \cite{Coverdale:2007a};
modeling suggests that the neutrons are thermonuclear  \cite{Coverdale:2007b}.

The configuration analyzed here is 
referred to as a $\lq\lq$Staged Z-pinch" and  
consists of  a high-atomic-number plasma liner 
imploding onto a solid-fill hydrogenic target: for example, Xe onto DT.
We have studied this configuration for some time. This specific name was 
applied  to characterize sequential, 
energy-transfer processes that occur
in these more complex load
configurations, leading to 
faster-risetime in the presence of 
improved stability.

As the current builds in the Staged Z-pinch
and the outer  liner begins to accelerate, shocks 
form, transporting  current and energy radially inward toward the target. 
As the  shock collides with the DT, a secondary shock is produced
in the target that also transports current and 
energy,  pre-heating the target. 
As the liner continues to accelerate and compress the target, 
a fusion burn begins in the presence of a 
flux-compressed, ultra-high magnetic field. The ultra-high magnetic
field confines fusion $\alpha$-particles, providing an additional source of 
heat for the  DT, raising its  temperature to 
50 keV and  causing ignition 
in  a magneto-inertial compression. For a 
precise set of initial-implosion parameters, 
net-fusion energy is produced. In the absence of  shocks 
the   radial-compression ratio needed  for fusion to occur,  
by adiabatic compression alone,  would be much higher. These are the
findings reported here.

The discussion proceeds as follows: 
Section II discusses the growth of RT instability and its effect on energy
coupling; Section III simulates the compression of a  Staged Z-pinch, 
for different initial parameters; Section IV describes the optimum
configuration for fusion break-even; and Section V concludes the paper.

\bigskip
\noindent
\begin{center}
{\bf II. Growth of the Rayleigh-Taylor instability}
\end{center}

A typical Z-pinch is Rayleigh-Taylor unstable during
implosion; since the implosion involves a light fluid (the magnetic field) 
accelerating  a heavy fluid (the plasma). In the linear-regime of analytic
modeling, plasma perturbations grow as \cite{Rahman:2004},
\begin{eqnarray}
\xi=\xi_0 e^{\gamma t}\;
\end{eqnarray}
where $\xi_0$ is the initial perturbation, $\gamma=\sqrt{gk}$ is the growth 
rate, $g$ is the acceleration,  $k$ is
the wave-number, and $t$ is the time. Approximating the
distance, $R$, over which the Z-pinch plasma is accelerated,  $R = gt^2/2$, 
Eqn. 1 may be  re-written as,
\begin{eqnarray}
\xi=\xi_0 e^{\sqrt{2Rk}}\;
\end{eqnarray}
and for a given mode number the
perturbation  growth  depends exponentially on the distance over which the plasma
is accelerated. Hence, Z pinch implosions from a small initial radius are preferred.

The accumulated Z-pinch energy (in Gaussian units) can be estimated roughly as 
the work done on the pinch, 
\begin{eqnarray}
W=\int{\vec {F} . {d\vec {r}}}={{I}^2 h\over c^2}\ln {[{R_i/R_f}]}
\end{eqnarray}
where $I$ is the current, $h$ is the axial length of the 
pinch, $R_i$ is the initial radius, and $R_f$ is the final radius of the pinch. 
Thus, high current and a  large radius implosions are preferred. The 
combined implications of Eqns. 2 \& 3 are that the  radius must be chosen 
judiciously to avoid instability, while accumulating high energy in the pinch. 

\newpage
%\bigskip
\noindent
\begin{center}
{\bf III. Simulations}
\end{center}

\begin{figure}[t]
\includegraphics{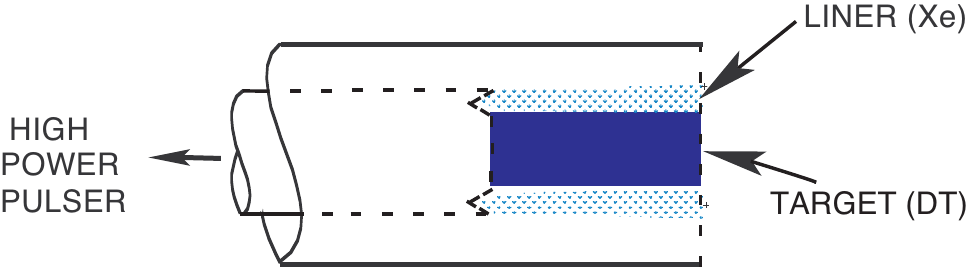}
\caption{(Color) Schematic illustration of the 
Staged Z-pinch.} 
\label{load}
\end{figure}

The simulated Z-pinch load configuration is a  1.5-cm long,  
0.2-cm thick xenon plasma liner imploding onto a 
DT  target, as shown in Figure \ref{load}. 
The initial Xe mass distribution is Gaussian  
and the DT target is uniformly-filled. The $\lq\lq$cold-start"
 initial plasma temperature is  2 eV, for both the Xe and DT. 

The implosion dynamics are simulated with
MACH2, which  is a single-fluid, 
magneto-hydrodynamic, 2-1/2 dimensional,
time-dependent code, 
that treats the electron, ion, and radiation
temperatures separately and calculates resistive and thermal diffusion
using established transport models \cite{Peterkin:1998}. 
The plasma equation-of-state is
determined from SESAME look-up tables 
(\url{http://t1web.lanl.gov/doc/SESAME_3Ddatabase_1992.html}). 
The Xe calculation uses the 
SESAME tables for thermal conductivity
and electrical resistivity.  A Spitzer model is used 
for the DT thermal conductivity and electrical
resistivity, since this data is not readily available 
in the SESAME tables. MACH2 
calculates flux-limited, single-group, implicit-radiation
diffusion. Ohm's Law includes the Hall Effect and 
thermal-source terms for magnetic fields. 

The plasma volume was resolved 
into 160 radial cells and 120 axial 
cells for 2-D simulations; sufficient to 
model  axial-instability wavelengths as small as 0.3 mm. 
A random seed perturbation of 0.01 (1$\%$) was
applied throughout the simulation volume. This seed value is  
arbitrary, yet is typical for  plasma liner 
simulations of this type. For comparison, a value 
of 25\% is  typically used for 
wire array simulations.  

The MACH2 code includes a self-consistent circuit model 
for the pulse-power driver  parameters (inductance, capacitance, 
and resistance) and the dynamically-computed pinch-plasma 
parameters (inductance and resistance). For our simulation 
the assumed short-circuit discharge parameters, resistance, inductance,
capacitance, and charging voltage were: 
R = 0, L = 10 nH and C = 0.35 $\mu$F, and V$_0$ = 3.4 MV. The 
capacitor-stored energy  is, 1/2 CV$_0^2$ =  2 MJ and the current-pulse 
 risetime is, $\tau_{1/4} \simeq$ 95 ns. The discharge 
electrodes are assumed to be perfectly conducting.

MACH2 was  run first in a 1-D mode,
for a fixed pinch-implosion time of 121 ns,
to determine the load masses, $M$, needed to implode from  the following 
initial radii: $R_i$ =  2.0, 1.5, 1.0, and 0.5 cm. For zero dimensional
modeling this is equivalent to keeping 
 $MR_i^2$ = constant for a given current  profile.
These 1-D mass parameters were then used as inputs for the 
2-D simulations. The respective mass 
densities for Xe  (order of decreasing radius, 2.0 - 0.5 cm) were: 
$\rho_{Xe} = 1.3\times 10^{-3}$,  $4.2\times 10^{-3}$, $1.7\times 10^{-2}$, and
$0.18$\ gm/cm$^3$. The respective  mass  densities for  DT 
were: $\rho_{DT} = 9.4\times 10^{-5}$, $5.5\times 10^{-5}$, $47.8\times 
10^{-5}$ and $3.4\times 10^{-3}$\ gm/cm$^3$. Note that the 
total mass of the DT is orders of magnitude smaller than for the Xe. 

\begin{figure}[t]
\includegraphics{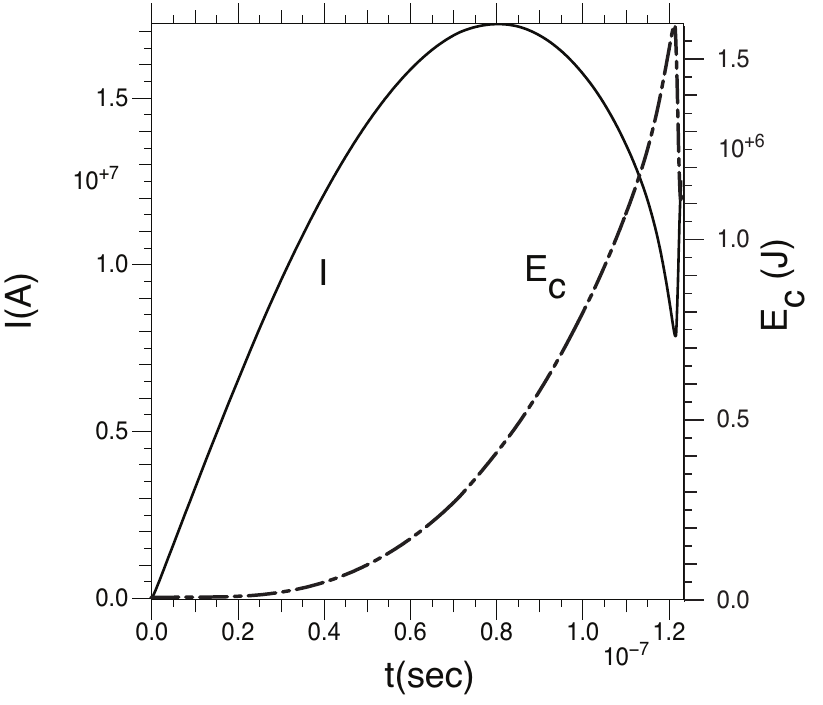}
\centering
\caption{Load current, I,  and the extracted circuit energy, 
E$_c$, for the 0.5 cm radius pinch.} 
\label{loadcurrent}
\end{figure}

Figure \ref{loadcurrent} plots the load-current time-profile,
I$_{load}$, for the 0.5-cm radius case. A peak current of I$_{load}$ = 17 MA is obtained in 80 ns.  
After the peak the slope of the I$_{load}$ waveform changes dramatically,  
due to the large increase in inductance caused by
the decreasing radius of the pinch; since 
L$_{pinch}(t) = \mu_0/2\pi~$ ln$[R_{out} / R_{pinch}(t)]$, where $R_{out}$ is  the radius 
of the  return-current boundary path
and $R_{pinch}(t)$ is the time-dependent pinch radius. 
Maximum compression of the Z-pinch occurs 
at 121.17 ns, when the  discharge current 
is  approximately I$_{load} \simeq$  8 MA. After this time 
the  pinch radius expands and the discharge  current 
rebounds. This time-dependent behavior of the current pulse is typical
for  all pinch radii that were simulated.

The time-profile 
for  the total energy extracted from the circuit, E$_c$,
is also displayed in Figure \ref{loadcurrent}, where 
E$_c$ =  E$_{radiation}$ + E$_{transport}$ + E$_{kinetic}$ + E$_{thermal}$ + E$_{inductive}$,
that is  the sum of the energy lost by dissipation (through radiation and transport), 
pinch energy  (kinetic and thermal),  and system energy 
(inductive magnetic). As illustrated 
E$_c$ continues to increase until
the moment of peak compression, reaching a peak value of, E$_c$ = 1.6 MJ.
The balance of the initial-stored  energy at this time remains in the circuit
capacitance and is equal to 0.4 MJ. 
After peak compression E$_c$
decreases rapidly,  indicating, as discussed below, that energy is added back
into the circuit as  the pinch explodes due to  the onset of fusion.

\begin{figure*}
\includegraphics{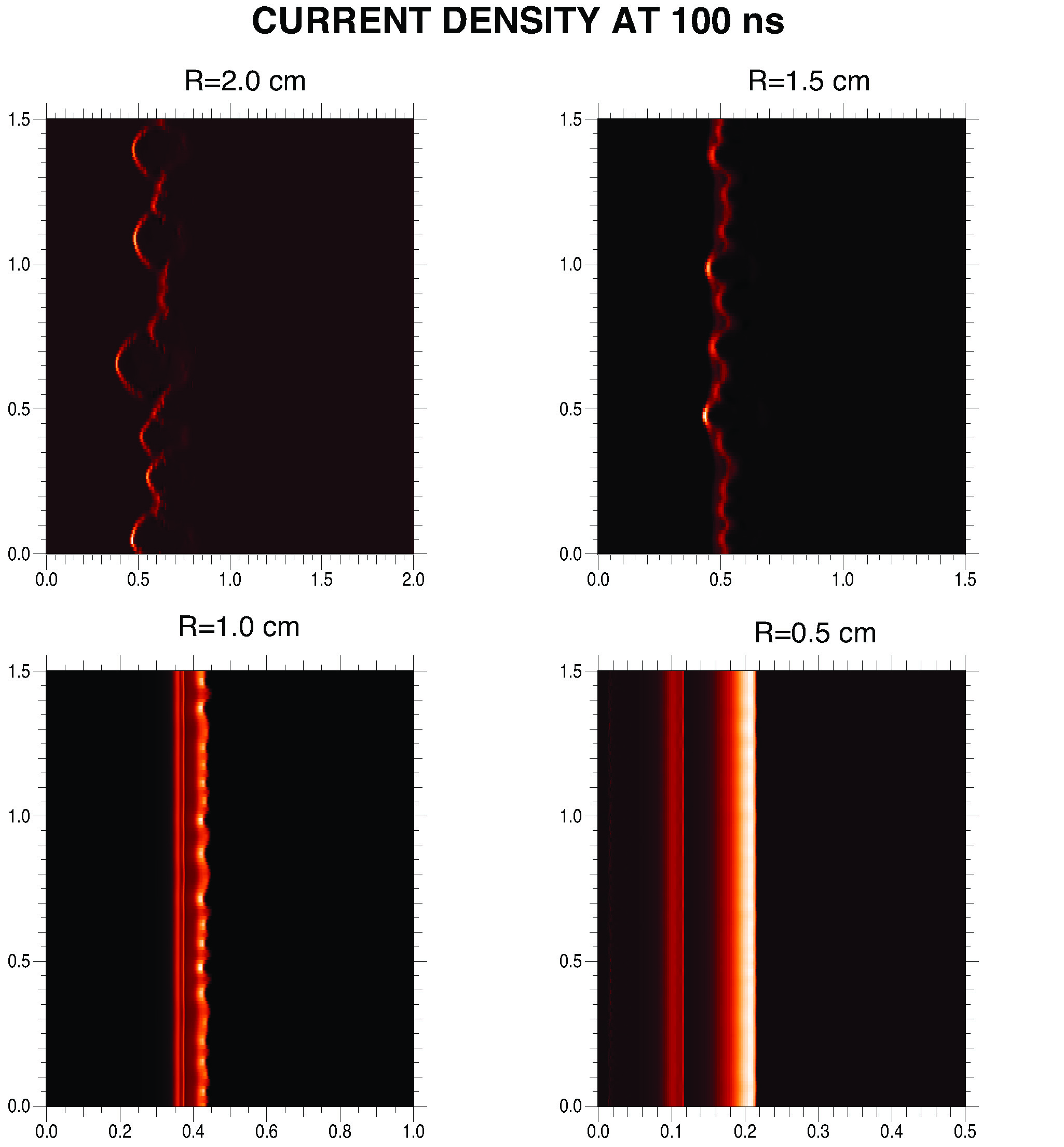}
\caption{(Color) Current density (R-Z) iso-contour profiles 
computed 100 ns into the
implosion as a function of the initial radius:  R$_i$ = 2.0 cm, 1.5 cm, 1.0 cm 
and 0.5 cm. The  horizontal and vertical axes
correspond to the radial and axial coordinates, 
respectively, and the
units are in cm. White color
corresponds to highest-computed intensity. }
\label{variousR_currentdensity}
\end{figure*}

Figure \ref{variousR_currentdensity} compares 
graphically the level of instability computed at 100 ns, 
as a function of the initial-pinch radius.
Shown are 2-D plots of  the pinch-current density.
In these illustrations the  R axis is horizontal and Z axis is vertical 
and the scale units are cm. 
In qualitative agreement with  Eqn. 2,  the level of instability
is greatest  for the  largest initial radius and smallest for the smallest 
initial radius.

For the 2.0 cm initial-radius simulation the calculation terminates at 101 ns, 
as the pinch becomes unstable and the discharge-current path is broken. 
The result is similar for the 1.5 cm initial-radius simulation, which terminates at 102 ns. 
For both the simulated neutron yield was insignificant.

For  the 1.0-cm radius simulation 
Figure \ref{variousR_currentdensity} shows
that the liner's outer surface is slightly unstable.
At the inner-radius edge of the  liner
is a stable  detached-current layer, illustrating how  a multi-liner cascade
can improve stability.
The 1.0 cm simulation terminated at 115 ns, due
to instability and the  neutron yield was
$Y = 2.7\times 10^{13}$. 

The  0.5 cm radius implosion provides the best stability, as
shown in Figure \ref{variousR_currentdensity}. At this 
time both the  liner and target surfaces remain stable.
The detached-current layer, evident in the preceding
panel, has broadened for this radius
and is located at the outer surface of the
DT target. The 0.5-cm radius case is now  examined in more detail.

\bigskip
\noindent
\begin{center}
{\bf IV. Detailed discussion for  the  0.5-cm radius implosion}
\end{center}

\begin{figure*}
\includegraphics{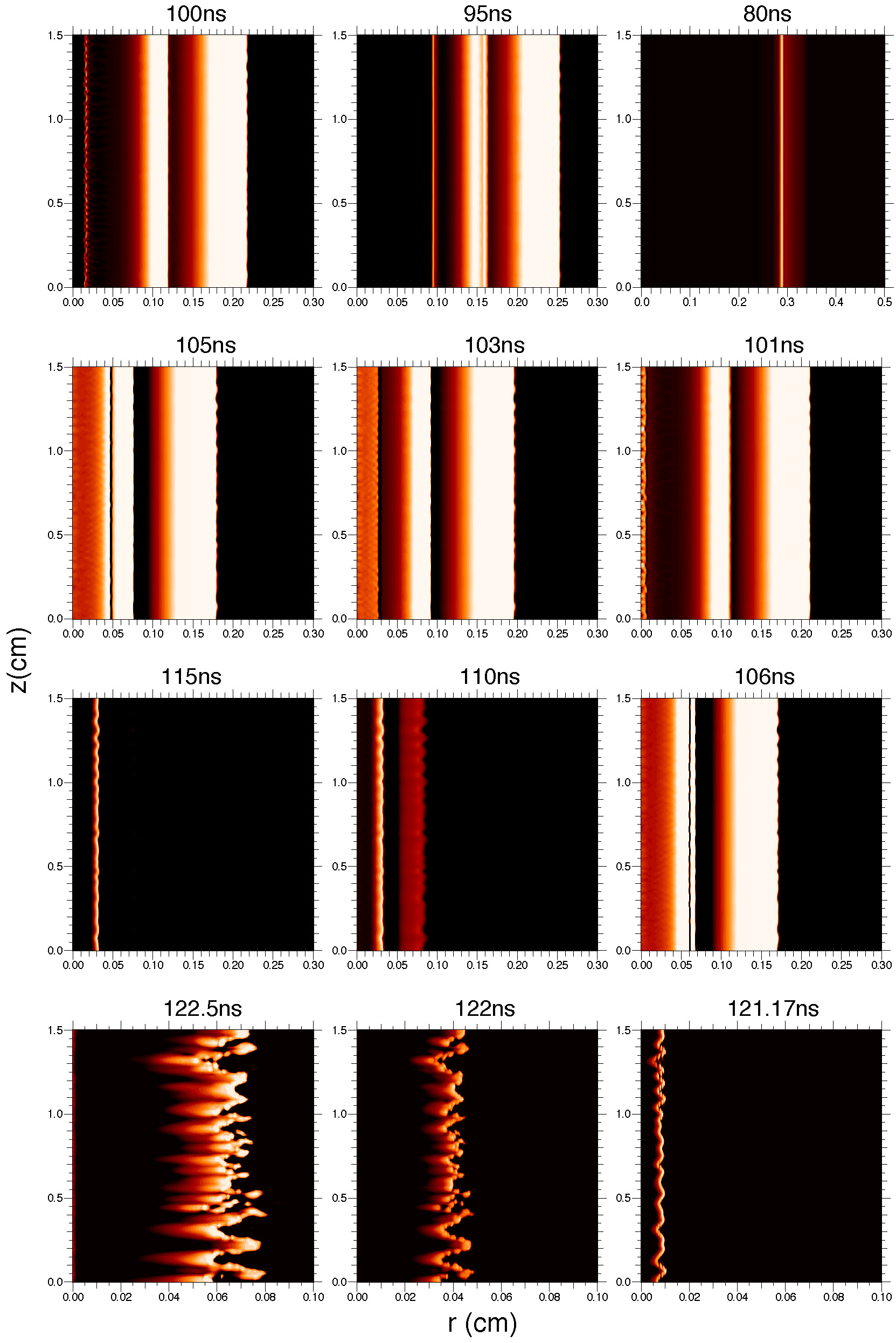}
\caption{\label{fig:wide}(Color) R-Z  profiles of axial current density 
for a 0.5 cm initial radius.}
\label{current}
\end{figure*}

Figure \ref{current} displays a time-sequence of   
2-D (iso-contour) discharge-current 
density images for the 0.5-cm radius implosion,  
beginning at 80 ns  and progressing
to 123 ns,  just after peak-compression of the pinch,
which occurs at 121.17 ns.  
Note  the radial scales in subsequent images: 
it is largest 
for  the 80 ns panel and shortest for the 
last three panels  (121.17, 122, and 122.5 ns).
These images  extend in time those that are
 shown in  Figure \ref{variousR_currentdensity},
illustrating the stability of  the 0.5-cm radius implosion
up to  peak compression, when the RT  instability develops  rapidly
and within a couple of ns,  the pinch disassembles. The RT instability
at the DT surface is due to deceleration of the liner against the internal
pressure of the target plasma and then subsequent expansion outward as the 
target plasma explodes.

\begin{figure*}
\includegraphics{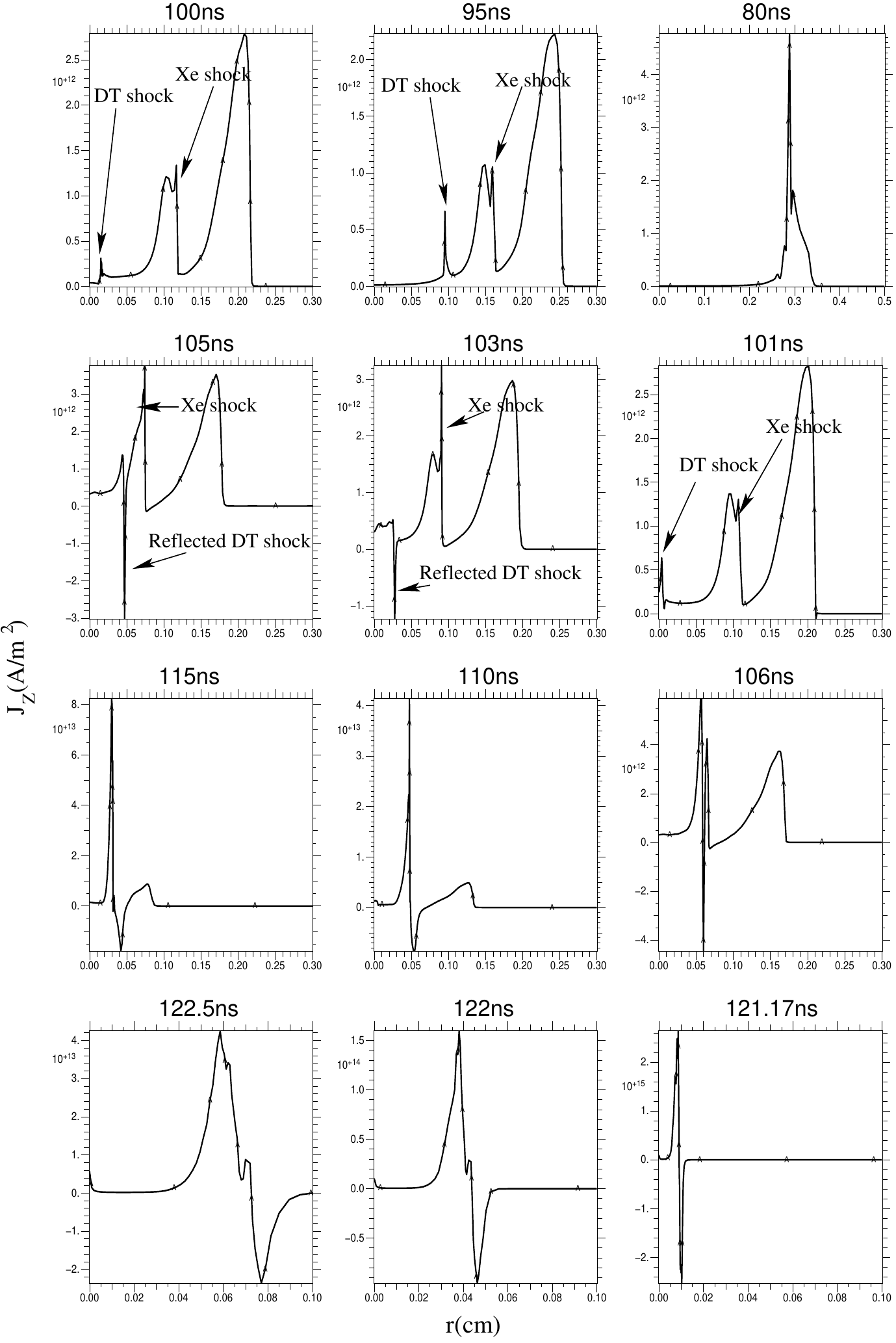}
\caption{\label{fig:wide}Axial current density
averaged over the axial direction.}
\label{current_1d}
\end{figure*}

An axial average of the data in Figure \ref{current} is 
also displayed  as line-plots in  Figure \ref{current_1d}.
At 80 ns  the outer radius of the pinch 
is at 0.34 cm, having
imploded from its initial
value of 0.5 cm. At this time the  current density 
is concentrated in a  narrow sheath  between 0.28 cm and 0.29 cm 
inside the DT and a more diffuse layer distributed 
between  0.3 cm and  0.34 cm radius inside the Xe. 
Until peak compression the value of the current 
density  at the outer surface of the Xe remains 
at a nominal value of, J$_z$ = 3-5 x 10$^{12}$ A/m$^2$
accelerating the outer surface of the Xe, by the 
$\vec J_z\times \vec B_{\theta}$ force. 
 
As time progresses, the current density is  
transported into the DT interior by shocks, which 
are labeled in the illustrations. Shocks develop
when the liner temperature
remains low and the density high \cite{DeGroot:1997b}. 
The specific choice of Xe, as a high Z liner material,
enhances the generation of shocks, by radiative cooling.

At 95 ns, when the outer  radius of the pinch has decreased to  0.25 cm, 
there are three layers of high current density:
an outer layer between 0.2 cm and 0.25 cm radius, a thinner layer  
between 0.12 cm and 0.15 cm, and a  third  layer  
at 0.09 cm. The Xe shock is characterized by a spike 
in J$_z$ that is located at the Xe-DT interface 
and a more diffuse region extending into  the DT. 
A second  shock is 
present inside the DT that is  generated when the Xe shock 
collides with the outer surface of the DT.

As time increases the DT shock converges to the  
axis and reflects. The outwardly propagating 
current layer splits into 
layers of positive and negative current density.
As the shocked layer expands outward,
 a self-consistent current loop is established inside the
pinch that supports flux compression.
The amplitude of the negative-current-density 
layer continues to grow 
as the liner implodes. 

At 121.17 ns the positive and
negative current densities have  
approximately equal values at, 
2.6 x 10$^{15}$ A/m$^2$. The radius of the
inner layer is  0.006 cm and the radius of the  outer layer is 0.011 cm. 
Until this time the magnitude and thickness of the current density
at the  outer surface of the  Xe liner remains relatively constant, 
when it begins to diminish and the implosion continues to be 
driven largely by liner inertia. 

Throughout the implosion there is a complicated evolution and interplay 
of the current density layers and their magnitudes. 
Even though the current densities are comparable in magnitude until
peak compression, the current-layer thicknesses are unequal.  
Integrating J$_z$ as a function of radius, confirms the presence of the
current loop. Near peak compression the
 magnitude of this loop current is,  I$_{loop}  \simeq 25 - 30$MA, 
while the total circuit current 
remains  as indicated in Figure \ref{loadcurrent}. 
The inductance of this current loop is approximately 0.7nH, with an 
equivalent inductive energy of 0.2-0.3 MJ. 

\begin{figure*}
\includegraphics{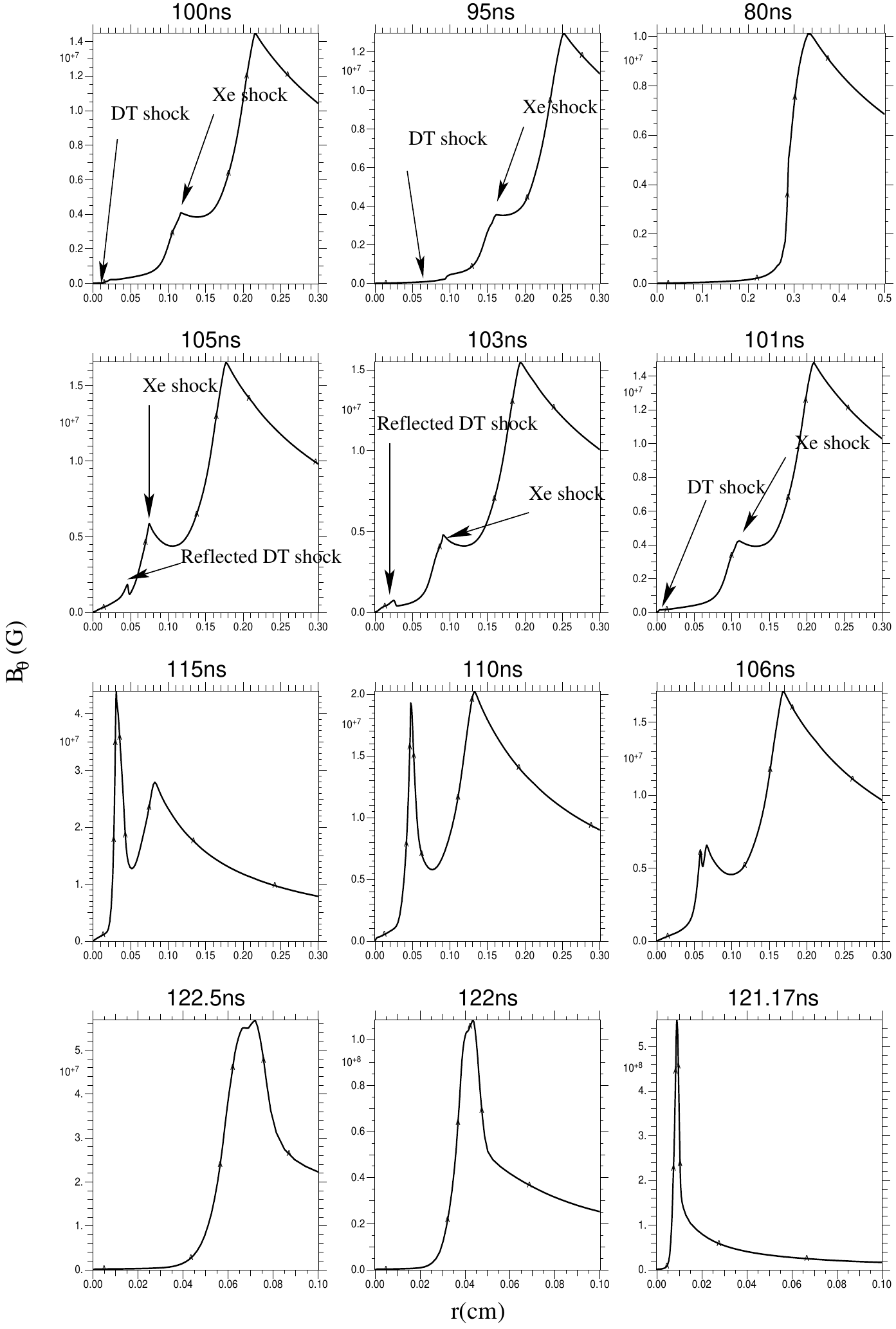}
\caption{\label{fig:wide} Azimuthal magnetic field
averaged over the axial direction.}
\label{magnetic_1d}
\end{figure*}

Examination of the corresponding  
azimuthal-magnetic  field, B$_{\theta}$, facilitates  
further insight. At 115 ns,  in Figure  \ref{magnetic_1d}, 
B$_{\theta}$ attains  a peak value 
of about 27 MG at the outer surface of the liner, whereas 
at  the surface of the DT,  B$_{\theta}$  has  
 a peak value of 44 MG. 
 It is this larger magnetic 
 field, located at the target surface, that  is 
 due  to flux compression, 
 driven  by  the inertial pressure of the liner  \cite{Rahman:1995}. 
 At 121.17 ns  B$_{\theta}$ attains a peak  value of  560 MG. 
At 122 ns the value of B$_{\theta}$ has decreased to  
100 MG as the pinch re-expands.

\begin{figure*}
\includegraphics{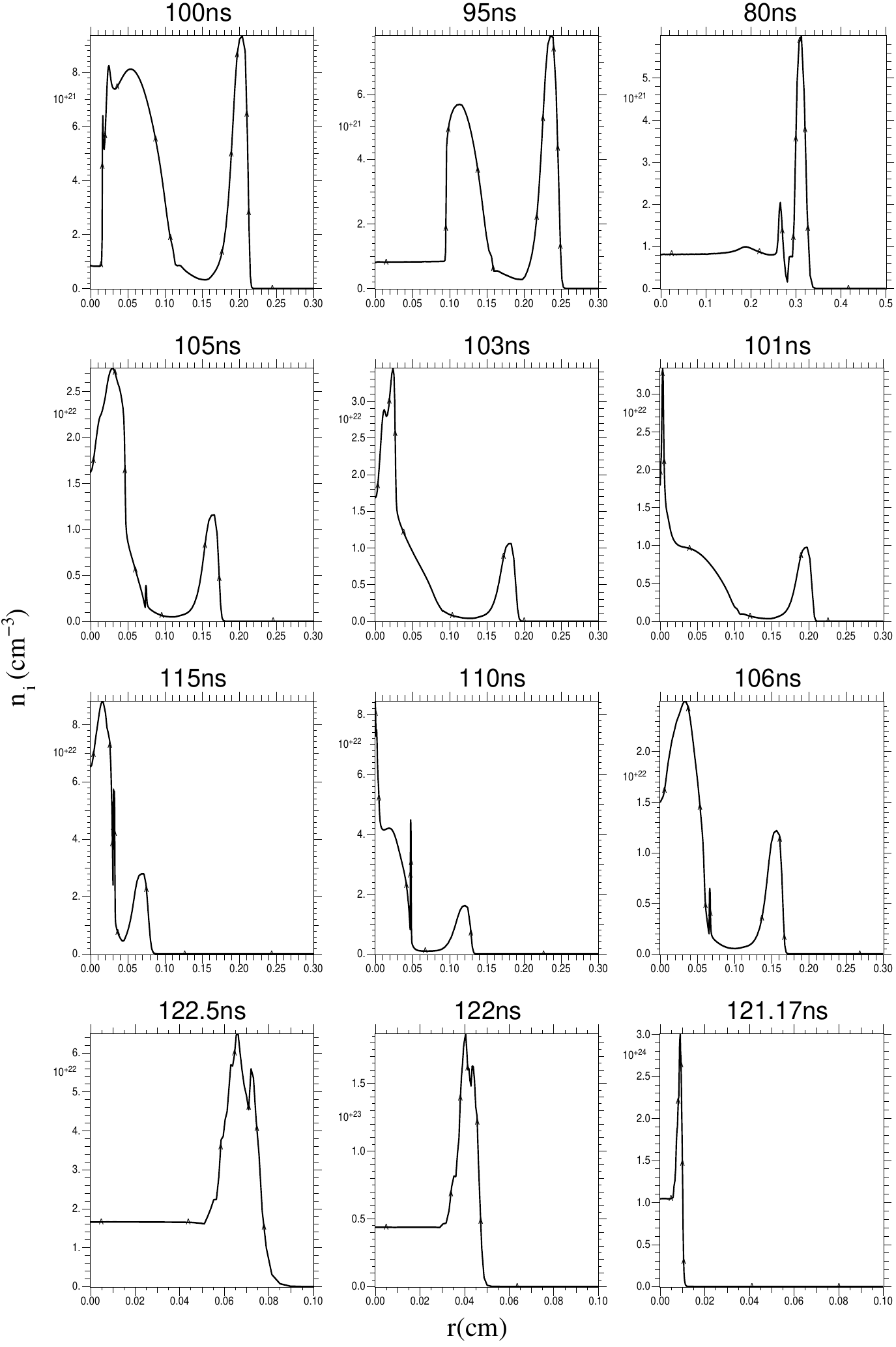}
\caption{\label{fig:wide}Ion density
averaged over the axial direction.}
\label{density_1d}
\end{figure*}

\begin{figure*}
\includegraphics{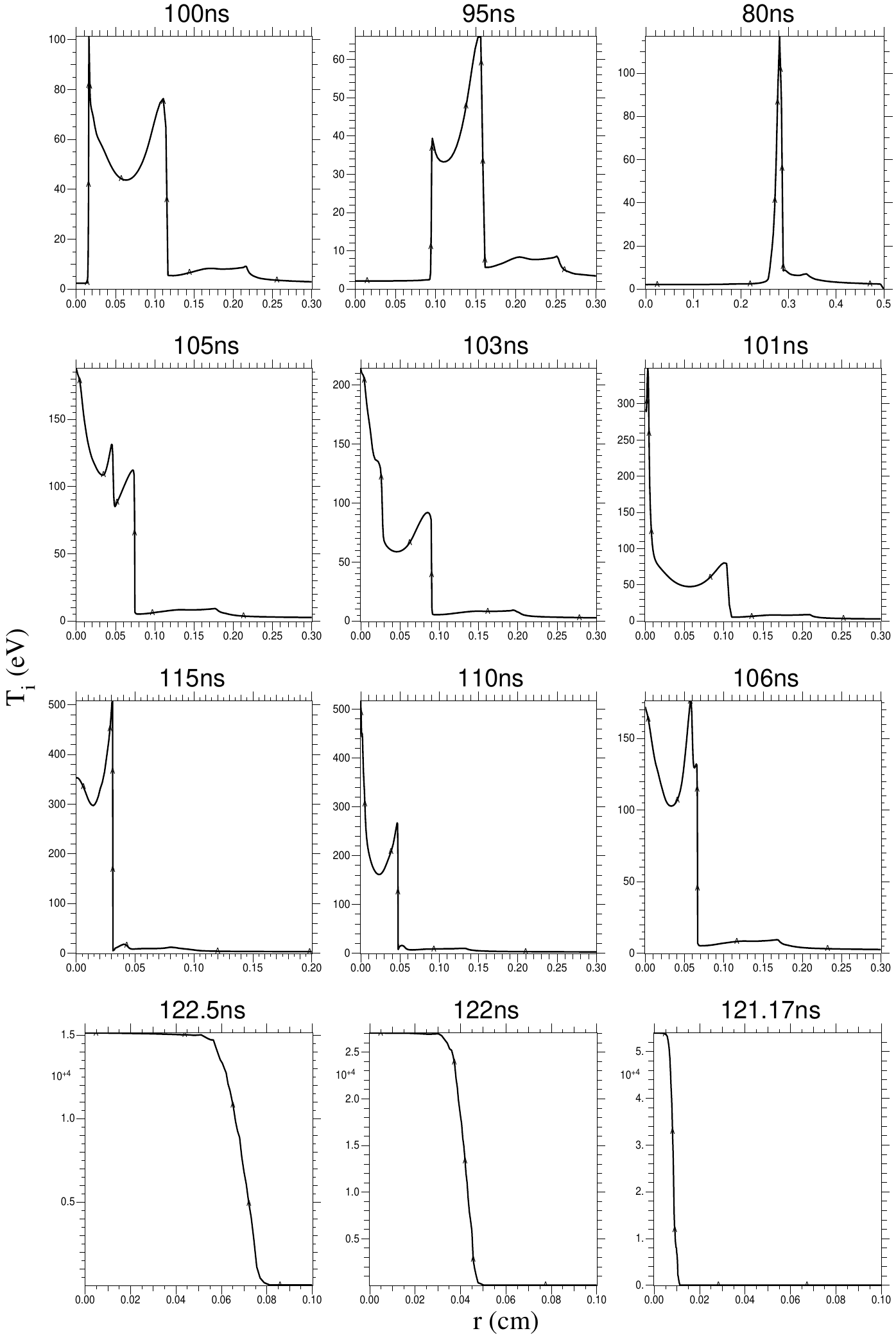}
\caption{\label{fig:wide}Ion  temperature
averaged over the axial direction.}
\label{temperature_1d}
\end{figure*}

The ion  density, n$_i$,  and temperature, T$_i$, 
panels are displayed in Figures 
\ref{density_1d} \& \ref{temperature_1d}, respectively.  
For all time steps  the electron and ion 
temperatures remain in thermal equilibrium, 
until  the moment of peak compression, 
when the fusion $\alpha$-particles heat the ions more rapidly than
the electrons.

The 80 ns panels show a sharp  transition in density and temperature
near 0.3 cm radius, with n$_i \simeq 6 \times 10^{21}$ cm$^{-3}$ and T$_i 
\simeq$ 115 eV. Notice that the peak density and temperature
do not coincide, as the higher temperature peak and lower density corresponds to the 
shocked DT, whereas  the higher density peak and lower temperature 
corresponds to the radiatively cooled Xe.
As time increases shocks continue to propagate in the
DT reflect and the density and temperature values oscillate. By 
115 ns the nominal  density and temperature
of the DT has reached,
 n$_i$ = 8 x 10$^{22}$ cm$^{-3}$,
and  T$_i$ =  400 eV. 
At peak compression,  121.17 ns,  the peak ion density in the DT is, 
n$_i$ =  10$^{24}$ cm$^{-3}$,
and the  temperature is, T$_i$ =  54 keV, whereas the peak ion density 
in the Xe is three times higher at, 
n$_i$ =  3x 10$^{24}$ cm$^{-3}$,
and the  temperature is much lower. The large
increase in temperature at this time is attributed to ion
heating by the fusion $\alpha$-particles, as will be discussed 
shortly. 

These  density and temperature values in the DT  are 
sufficient for fusion to occur. Based on  the Lawson criterion, 
n$\tau > 10^{14}$ cm$^{-3}$-s,  a confinement time of 
$\tau$ = 30 ps is needed, which is well satisfied. 

Shocks play a key role in the 
Staged Z-pinch, pre-compressing  target
plasma prior to the onset of  fusion. In the 
present case the DT is compressed from 
0.3 cm radius to a final radius of 0.005 cm, that is a factor of 60, 
c.f., Figure \ref{density_1d} \& \ref{temperature_1d}, producing a temperature 
increase from 2 eV to 60 keV. 

To achieve the same temperature increase in the DT without 
shock heating,  that is  by adiabatic compression alone, 
the radial compression ratio of the DT would need to be much higher.
For adiabatic compression the 
initial and final  plasma temperatures are related to the
initial and final radii by, T$_f$ = T$_i(R_i/R_f)^{2(\gamma - 1)}$, where 
$\gamma$ = 5/3 is the
ratio of specific heats  for DT.
Hence, to achieve the same temperature increase noted above,
the compression ratio would need to be, $R_i/R_f \simeq$ 2.5 x 10$^3$.
Thus,  shock heating is a important
component in  preparing the DT target for fusion conditions.

\begin{figure*}
\includegraphics{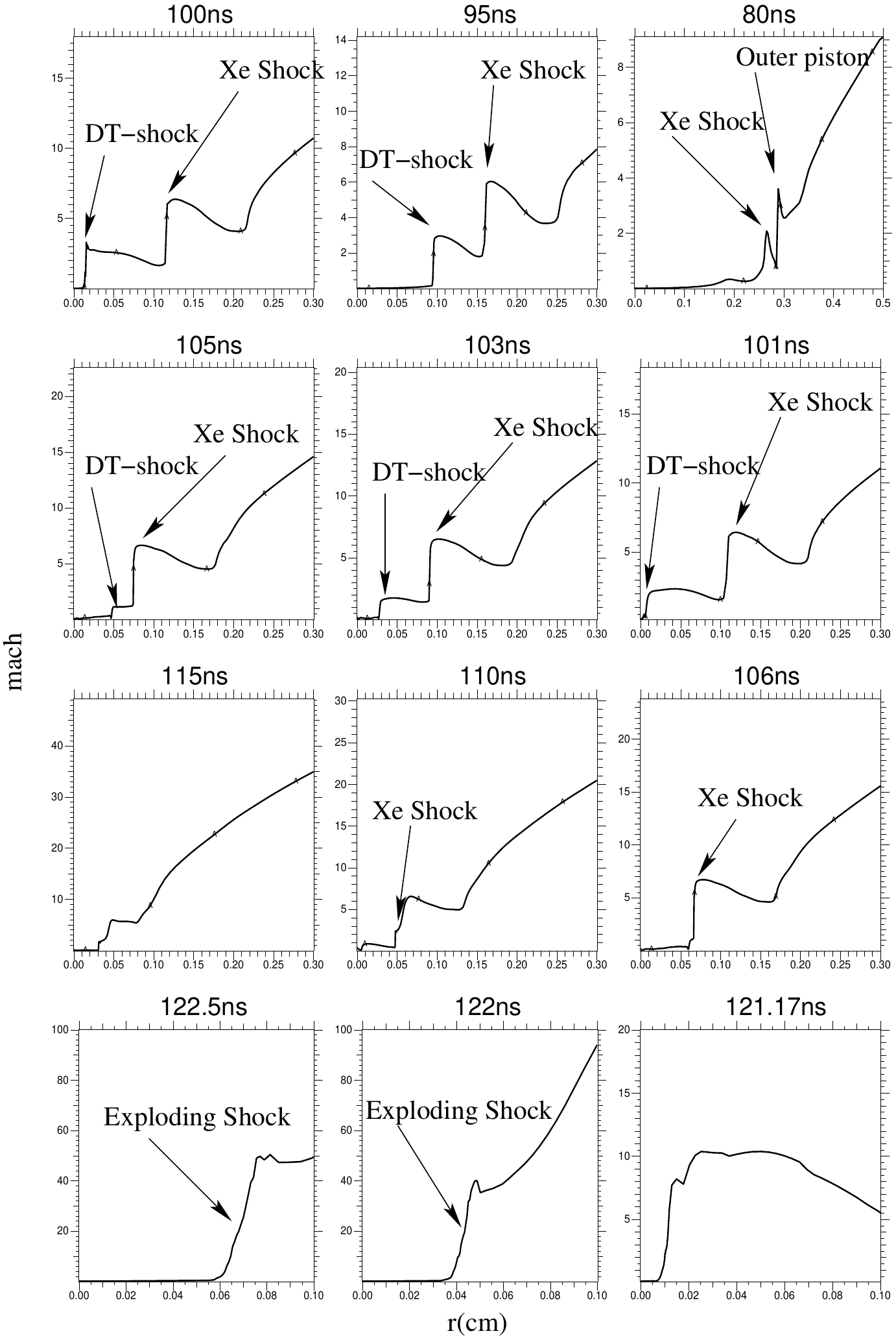}
\caption{\label{fig:wide}$\lq\lq$Line-outs" of mach number 
averaged over the axial direction.}
\label{mach_1d}
\end{figure*}

Consider 
the Mach number, $M\, =\, V_r/C_s$,
where $V_r$ is the radial implosion velocity and 
$C^2_s= \partial P/\partial \rho$ is the sound speed, $P$ is the pressure
and $\rho$ is the density. 
The Mach number 
was calculated at each simulation-grid point
based on this equation;  
the results are displayed in  Figure \ref{mach_1d}.

At 80 ns a Mach 2.2 shock is present at 0.27 cm radius, just  
inside the DT.
At 95 ns, when $V_r$ =  4 cm/$\mu$s, 
two shocks are clearly evident: one located at 
the Xe-DT interface and another
located inside the DT, at 0.093 cm radius, 
confirming the labels shown in preceding  Figures.
The lower Mach number for DT 
is due to the high sound  speed in DT.
During the next few ns the DT shock 
diminishes in amplitude, as it converges to the 
pinch axis, reflects, and then recollides with the 
imploding Xe shock  at 106 ns.  

At peak compression 
the DT ignites, and a radially expanding 
shock is generated on axis. 
At this time the  pinch outer radius  is, 
$R_f$ = 0.01 cm and 
the azimuthal magnetic field is, B$_\theta$ = 560 MG.
Such a high-intensity magnetic field traps the
fusion $\alpha$-particles inside the target, providing an
additional source of heating for target ignition. 
When  the  DT plasma
pressure  exceeds the inertial  pressure
of the liner,  the pinch explodes, driven
by the production of internal  energy.

Some flux leakage occurs, but we assume
that most of it remains during the  fusion burn.
Assuming an average value for B$_\theta$ of 100 MG and
a 3.5 MeV $\alpha$-particle energy, 
the corresponding  
$\alpha$-particle gyroradius would be, 
$\rho_{\alpha}$ = 0.002 cm, 
which is roughly an order of magnitude less than the
final pinch radius, $R_f$. 

Based on the 1-D line averages presented above, 
nearly all of the initial Xe mass 
participates in the implosion and
is  compressed into a thin layer, approximately 
0.001 cm thick,  at the outer surface of the DT,
at peak implosion, with an average radius of 0.01 cm.
Although some instability  is evident at this time, with
an approximate wavelength of $\lambda$ = 1 mm,
the effect of instability is exaggerated by the 
expanded radial-scale displayed here,  
relative to the axial-scale length. For the most part the pinch 
remains largely intact and stable as the fusion-burn proceeds.  

\begin{figure*}
\includegraphics{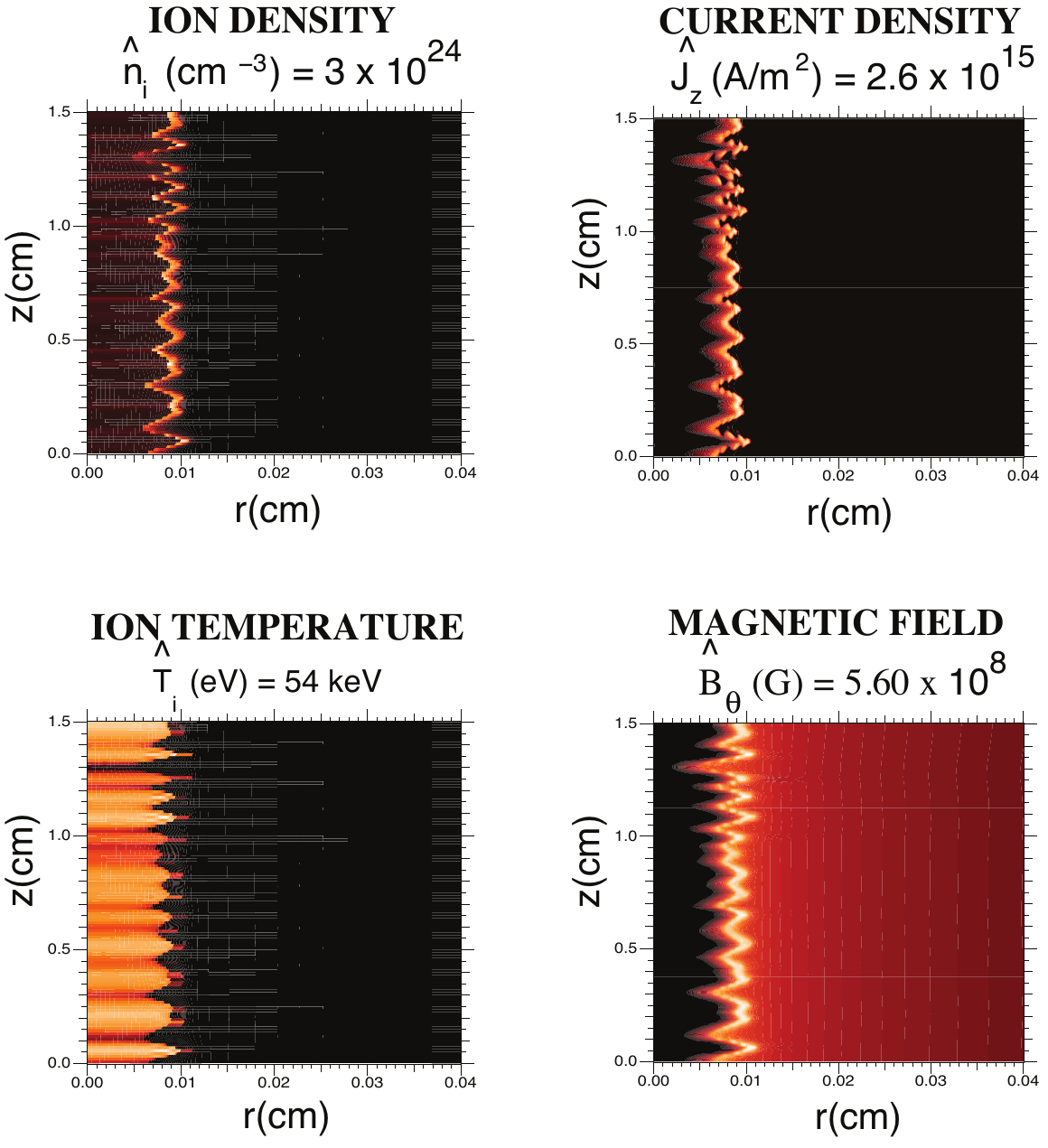}
\caption{\label{fig:wide}(Color) R-Z iso-contour profiles of (left to right, top to bottom) ion 
density, axial-current density, ion temperature,  and azimuthal-magnetic 
field computed 121.17 ns into the compression. The peak-parameter values 
are shown at the top each panel, as $\hat n_i, \hat J_z, \hat T_i, \hat B_{\theta}$.}
\label{peak_par2}
\end{figure*}

Figure \ref{peak_par2} provides a 
summary of the  simulated pinch parameters
at 121.17 ns
(ion density, ion temperature, axial-current density, 
and magnetic field). Also indicated are the 
peak values for each parameter, 
e.g., $\hat n_i, \hat J_z, \hat T_i, \hat B_{\theta}$, obtained from the 
1-D line-outs, above. Note the magnified radial 
scale of these images, relative to previous figures. 

The illustrations in Figure \ref{peak_par2}  provide a 
good indication of the final radial-compression 
ratio for the pinch at peak implosion. Taking the initial radius of,
$R_i$ = 0.5 cm and the final radius of, $R_f$ = 0.0067 - 0.011 cm, 
the pinch compression ratio is in the range, $R_i/R_f \simeq$ 45-75. 
Radial-compression ratios  of 40-45 have been reported 
in an experimental 
configuration similar to the one analyzed here;  
that is for a multi-shell, gas-puff implosion
 \cite{Failor:2007, Chaikovsky:1997}. At the higher end, 
a radial-compression ratio in excess of
120 has been reported for an extruded-shell   
Z-pinch \cite{Appartaim:1998}. So  the  compression
ratio reported here is reasonable.

As shown in Figure \ref{peak_par2}, 
less than a ns after peak compression, the entire Z-pinch 
column becomes RT unstable and the pinch
disintegrates (light-fluid DT plasma pushing against the
heavy-fluid Xe liner). The
last recorded output for data is at 122.5 ns, and 
the calculation stops at 123 ns. 

\begin{figure}
\includegraphics[scale=0.9]{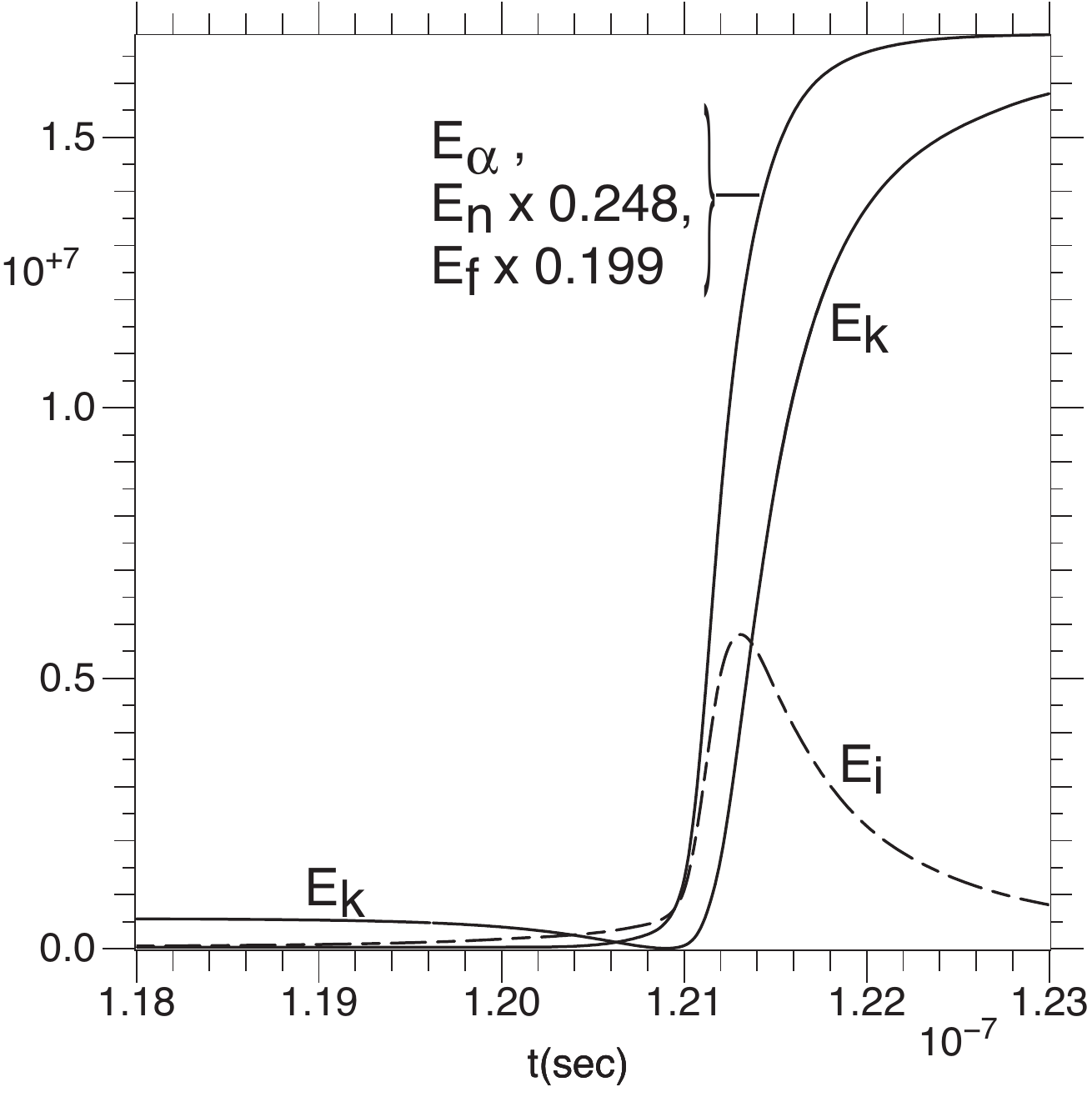}
\caption{\label{fig:epsart} Energies: alpha-particle energy, E$_{\alpha}$, liner-kinetic energy, E$_k$,
ion-thermal energy, E$_i$, scaled neutron energy, E$_n$ x 0.248, and scaled
total-fusion energy, E$_f$ x 0.199, plotted as
a  function of time.} 
\label{energies}
\end{figure}

The time-evolution of the Staged Z-pinch 
energies are shown in 
Figure \ref{energies} for the expanded 
time interval of, t = 118-123 ns, 
including: 
alpha-particle energy, E$_{\alpha}$, implosion-kinetic energy, 
E$_k$,   ion energy, E$_i$, scaled neutron energy, E$_n$ x 0.248, and 
scaled fusion energy, E$_f$ x 0.199. 

Near peak compression  E$_k$ decreases
to zero, from a  peak value of, E$_k$ = 0.06 MJ at 118 ns
as implosion kinetic energy is converted into plasma thermal energy. 
The  ion energy begins to increase  around
119 ns. At  120.9 ns E$_i$ begins to  increase rapidly to a  peak value of, 
E$_i$ = 0.6 MJ at  121.3 ns. The rapid  increase in E$_i$
at 120.9 ns  is driven by compressional and  $\alpha$-particle  heating. 
The pinch stagnation time is roughly  
given by the time interval between when E$_i$ increases rapidly
and E$_k$ remains low. From  Figure \ref{energies} t$_{stagnation} \simeq$  
0.3 ns. Near the end of the fusion burn,
at 121.2 ns, E$_k$ begins to increase rapidly, as the Z-pinch internal
pressure increases, rising to a peak value of 15.8 MJ 
at the end of the calculation. 
 
The total fusion energy is equal to the sum of the neutron and alpha
particle energies, E$_f$ = E$_{\alpha}$ + E$_n$, 
which attain their peak values at the end of the calculation, 
E$_f$ = 84 MJ, E$_{\alpha}$ = 17 MJ, and E$_n$ = 67 MJ
and a  neutron yield  of,  
$Y \simeq 3.0\times 10^{19}$.  
This level of fusion energy is 42 times  greater 
than  the energy initially stored the circuit
capacitor. For comparison, our 1-D simulations for the same 
configuration predict a 70 MJ yield.

\begin{figure}
\includegraphics{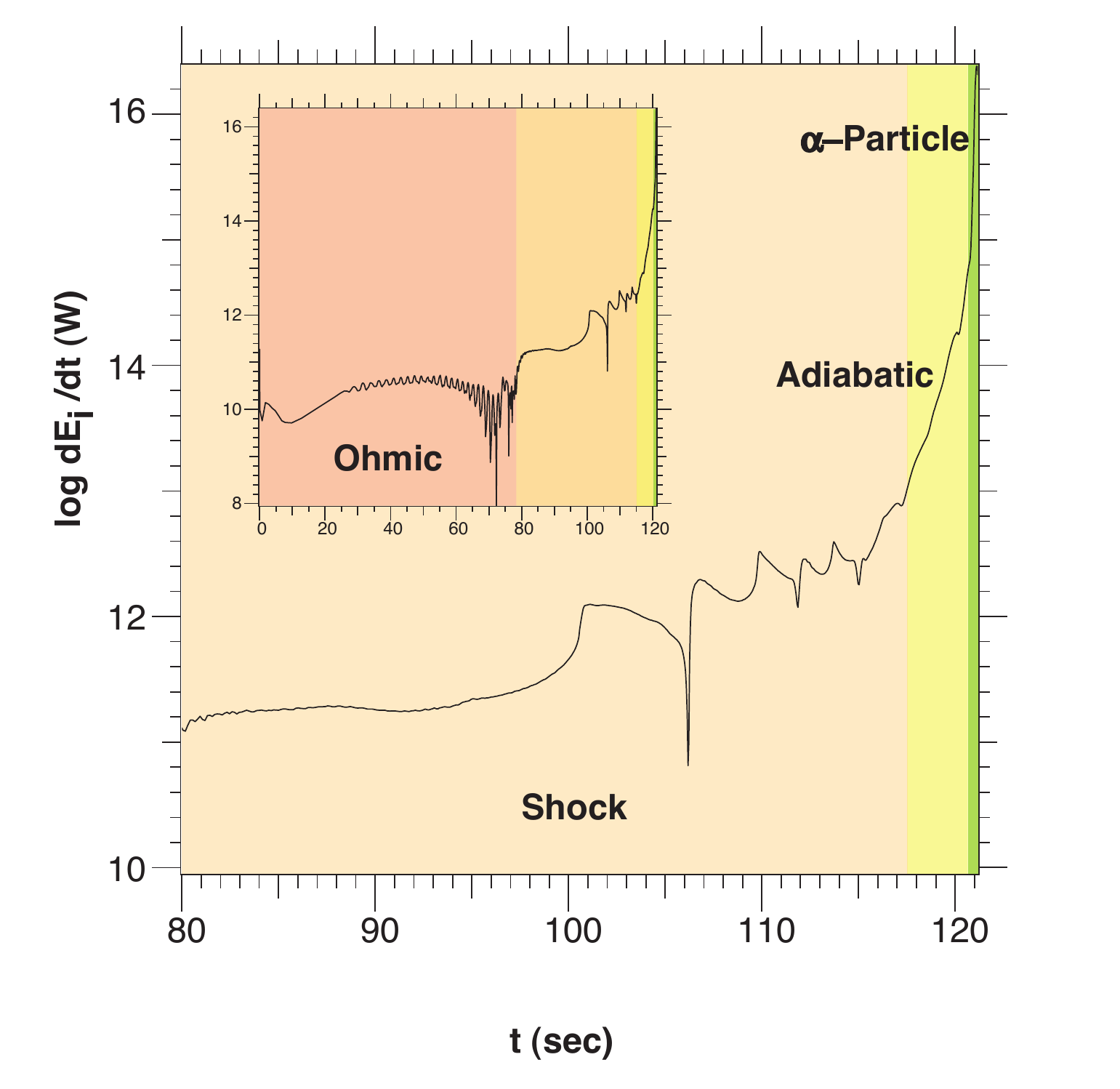}
\caption{\label{fig:epsart} (Color) Log total ion heating power, P$_i$ = dE$_i$/dt,  plotted as a  function of time.  The inset shows the full simulation time, from 0 - 123 ns. } 
\label{power}
\end{figure}

Fig \ref{power} shows the total 
ion heating power P$_i$ = dE$_i$/dt versus time, on a 
logarithmic scale. The full simulation time is shown in the inset. 
Four distinct time phases are evident, 
characterizing  the implosion 
dynamics of the Staged  Z-pinch. These phases are
defined  as the times when various ion-heating
mechanisms dominate and are labeled as: 
Ohmic, Shock, Adiabatic, and  $\alpha$-Particle. 

The value of P$_i$ displayed is due to 
Ohmic heating, which 
dominates in the early phase of
the implosion, from 0 - 78 ns.
This is the phase when  the $\vec J_z\times \vec B_{\theta}$ force 
remains small and the Z-pinch  plasma remains 
at rest.  As shown there is a competition between the
plasma pressure outward  and the magnetic  pressure inward,
leading to a quasi-static equilibrium, where P$_i$ oscillates \cite{Felber:1982}.

The  shock heating phase begins at approximately 80 ns,
as correlated with 
the density and temperature spikes shown in  Figures \ref{density_1d} 
\& \ref{temperature_1d}. During this phase a shock
propagates  toward and reflects off  the pinch axis, 
rapidly heating the target at 
one to two orders of magnitude higher power level  than
during the Ohmic phase. The oscillation observed during the
Ohmic phase, being several 
orders of magnitude less, is not
 observed during this and subsequent phases.

At 101 ns, the shock reflects off the axis and the
target plasma temperature is about 150 eV. 
From 101 ns to 121 ns shocks continue to reflect 
back and forth inside the target plasma
contributing further to ion heating.

From approximately 115 ns  onward,  adiabatic 
heating dominates until approximately 120 ns.
The onset of  $\alpha$-particle  heating occurs during 
the final ns, or so, when the target temperature
increases rapidly to approximately 50 keV and the plasma 
releases the maximum amount of fusion energy.

\bigskip
\noindent
\begin{center}
{\bf V. Conclusions}
\end{center}

This paper presents 2-D simulations of
a Z-pinch. The load is a 0.2 cm thick shell of 
Xe imploding onto a DT target (Staged Z-pinch). The 
system is driven by a 95 ns
risetime, 17 MA, 2 MJ pulser. 
The 2-D simulations were performed using  MACH2, a sophisticated
radiation-hydrodynamics code. 

We have considered several cases for the 
pinch initial radius, ranging from 2.0 cm down to 0.5 cm. 
The implosion dynamics are very sensitive to the 
choice of the initial radius, the atomic composition of the liner
mass, and the final pinch stability.  

High performance is obtained through careful
optimization of these parameters: i.e., the liner thickness, 
the liner mass distribution, and the target mass. 
The best stability is produced for the smallest initial
radius simulated. The pinch produces 
precisely timed shocks that 
originate in the Xe liner and are transmitted
across the mass boundary, into the DT target. 
Target shocks reflect inside the DT
as it is compressed by the Xe liner. 
For the duration of the  implosion
the pinch remains stable. 
There exists a theoretical basis 
to account for  enhanced stability in 
a shock-compressed implosion 
system \cite{Rostoker:1978, DeGroot:1997b}

The dwell time of the pinch at maximum compression 
is around 0.3  ns, as the pinch is compressed by the liner
inertia, in the presence of an intense azimuthal 
magnetic field; the latter is a result of flux compression. 
The implosion is accurately characterized as 
$\lq\lq$magneto-inertial" and produces a 
nominal radial-compression 
ratio for the liner of 50 and  a fusion-energy gain 
that is  42 times greater than the stored, 
capacitor-bank energy. 

3-D simulations of Z-pinchs  have 
appeared recently,  directed principally
toward the  analysis of  wire-array 
implosions \cite{Chittenden:2007, Yu:2008}. 
For wire-array loads a 3-D simulation is critical, since the
discrete nature of the wire array load inevitably 
introduces azimuthal non-uniformities. 
However, for the small initial radius of a Staged Z-pinch, 
and for a uniform, solid-fill liner,  a 3-D simulation
is not expected to be as critical.  Indeed, 
even for our 1-D calculations the predicted fusion-energy yield  
was  70 MJ, which is less than  2-D predictions for  85 MJ, where
one would expect a lower yield because of the higher dimensionality.
The higher yield  for the 2-D simulation is probably due to the 
appearance of hot spots, generated at the first collapse of 
the  on-axis shock. 

\begin{acknowledgments}

This project was supported by the US Department of Energy. 
We acknowledge R. E. Peterkin and J. H. Degnan for their assistance in
providing access to the MACH2 code.
\end{acknowledgments}

\bibliography{bibliography}
\bibliographystyle{unsrt} 

\end{document}